\documentclass[preprint,english]{elsarticle}
\usepackage{graphicx}
\usepackage{amsmath}
\usepackage{amssymb} 
\usepackage{epsfig} 

\begin{document}

\title{Unfolding of phases and multicritical points 
in the 
Classical 
Anisotropic van Hemmen Spin Glass Model with Random Field}

\author{S. G. Magalhaes}
\ead{sgmagal@gmail.com} 
\author{I. C. Berger, R. Erichsen Jr.}
\address{Instituto de F\'{\i}sica,
  Universidade Federal do Rio Grande do Sul, Porto Alegre, RS, Brazil}

\date{\today}

\begin{abstract}
 
We study magnetic properties of the 3-state spin ($S_{i}=0$ and $\pm
1$) spin glass (SG) van Hemmen model with ferromagnetic interaction
$J_0$ under a random field (RF). The RF follows a bimodal distribution
The combined effect of the crystal field $D$ and the special type of
on-site random interaction of the van Hemmen model engenders the
unfolding of the SG phases for strong enough RF, i. e., instead of one
SG phase, we found two SG phases.  Moreover, as $J_0$ is finite, there
is also the unfolding of the mixed phase (with the SG order parameter
and the spontaneous magnetization simultaneously finite) in four
distinct phases. The emergence of these new phases separated by first
and second order line transitions produces a multiplication of triple
and multicritical points.

\end{abstract}

\maketitle

\section{Introduction}
\label{Sec1}

Disorder effects in magnetic systems with localized spins considering
random interaction, frustration and random fields are permanent
sources of challenging issues.  The simultaneous combination of these
complex examples of disorder does exist in real systems.
It can be found in Fe$_x$Zn$_{1-x}$F$_2$ and Fe$_x$Mg$_{1-x}$Cl$_2$
compounds \cite{Belanger98}.  More recently, spin glass and RF have
been also suggested to exist in the diluted Ising-like dipolar
ferromagnetic compound LiHo$_x$Y$_{1-x}$F$_4$ \cite{Gingras11}.  The
existence of real systems with the combined presence of a spin glass
(SG) phase and random field (RF) makes its description a quite
relevant problem in the disordered magnetism with localized spins
(see, for instance, Ref.  \cite{SNA}).

Such description has several obstacles.  One of them concerns the use
of replica method.  In particular, the dispute about the existence of
a replica symmetry breaking (RSB) in a Ising SG model as predicted by
Parisi \cite{Parisi80a}.  The crucial validation of this scenario
would be the existence of the de Almeida-Thouless line
\cite{Almeida78}. This existence is still currently disputed by
simulational techniques or even experimentally \cite{Mydosh2015}.  In
addition, for 3-state disordered spins models, the replica method presents
others difficulties.  These models are known by displaying
tricritical points and, therefore, a first order line transition whose
location is a complicated task within the replica method.  For
example,
it is known that in the Gathak-Sherrington model at mean field level
\cite{gs}, the proper location of the first order line transition is
far from
obvious since the stability requirements do not provide proper
guidance to select correctly the SG solution \cite{Costa94}.  Indeed,
the presence of a RF complicates the situation since the
first order phase transition line is much affected by the RF
\cite{Vaz2012,Vaz2013}.  Another aspect that should be remarked is
that the RF field induces the replica symmetric SG order parameter
which becomes finite at any temperature.  As a consequence, it is
strictly necessary to find RSB SG solutions of the order parameters
\cite{Morais2016} which makes even more complicated the location of
the first order line transition.  That situation enforces the
necessity to work with disordered spin models in the presence of RF
which avoids the replica method.

We present in this work an analysis of the combined effect of random
couplings and random fields in the 3-state spin ($S_{i}=0$ and $\pm
1$) van Hemmen model with a crystal field $D$, the so called
anisotropic van Hemmen spin glass (AvHSG) model \cite{vh1,vh2}. This
on site-disorder model was introduced as an exactly solvable model
suitable to describe metallic SG without the use of the replica method.
It has both fundamental ingredients for the SG phase, i. e.,
disorder and frustration.  Although its infinite-range version does
not present a complex free energy landscape with a large number of
local minima \cite{Choy1983}, the van Hemmen model can account for
several static properties of real SG systems.  Another important
aspect is that this model displays naturally mixed phases if a
ferromagnetic interaction is also present. In this mixed phase the SG
order parameter and the spontaneous magnetization are simultaneously
finite. Its existence is deeply related to the special type of on-site
random interaction of the van Hemmen model which are given as a
product of local random variables \cite{vh1,vh2}.
In the AvHSG model the number of sites with spins magnetically actives
can no longer coincides with the number of site lattices $N$.  

The combined 
effects of $D$
and the uniform field $h$ are well illustrated in
a previous study of the AvHGS model performed by de Almeida and
Moreira 
\cite{Almeida1986}. 
At $T=0$, for small $D$ and $h$, these authors found the usual spin glass phase with spin 
glass order 
parameter $q=1/2$ and magnetization $m=0$, as expected. Interestingly, 
for 
large
$h$ and $D$  frustration is still favored.  
As consequence, there is the
emergence of a new special type of mixed phase with 
$m=1/2$ (induced by $h$) and SG order parameter $q=1/4$. 
Moreover, at $T=0$, the phase transition lines between paramagnetism 
(PM), SG and this mixed phase are all first order. 
At finite temperature, the phase transition lines are first and second order. As 
consequence, there are  ordered critical and tricritical points.

The scenario described previously arises the question which are the 
consequences if the uniform field is replaced by a RF?  
While for small RF and $D$, one can expect the presence of the usual SG phase (as found in 
the case with uniform field), it is far from obvious what is the scenario when  RF and 
$D$ are large.
As result, we anticipate that the combination of effects coming from  RF and $D$ is 
responsible by  the unfolding of the SG phase in two. 
To be precise, at $T=0$, we find two solutions for
the SG order parameter instead of one, i. e., the usual one for small  RF and $D$  and a 
new one for large RF and $D$.
The phases corresponding to
each solution have genuine phase transitions between them.  Furthermore,
since we also analyse  the AvHSG model with a presence of a
ferromagnetic interaction, one might expect that the 
unfolding happens not only
with SG phase, but also with mixed phase. As consequence, one can also expect a 
complicated scenario of reentrant transitions between paramagnetic, SG and mixed phases.  
At $T\neq 0$,  the transitions between these several phases can be not only  first order 
but also second. Thus, one can have the multiplication of 
multicritical points. 

It should be remarked the role of RF or $D$ when acting alone. 
For instance, It  is well known that $D>0$ tends to destroy any magnetic long range order  
by favoring 
states $S=0$.  In its turn, previous results on the Ising van Hemmen model with RF  
\cite{Nogueira}  show that the RF  tends to suppress the SG phase. Therefore, it is quite 
clear that neither the RF nor the D acting 
alone can produce the unfolding  of phase obtained 
in our work. Lastly,  it is worth mentioning a recent work 
by Morais et al \cite{Morais} using a random crystal field (instead of a RF) following a 
bimodal distribution with a $p$ fraction of the spins without the 
influence of the crystal field and $p-1$ fraction under the influence of $D$. Their 
results show that when $p$ increases, the SG phase appears even for large $D$.

  This paper is structured as follows. In Section II, we discuss the
model and give details of the analytic calculations in the anisotropic
van Hemmen model with RF in order to obtain the order parameters. Our
numerical solutions for the order parameters are summarized in phase
diagrams which are displayed in Section III. The conclusion is
presented in Section IV.

\section{Model}

The model is given by the Hamiltonian

\begin{equation}
  H=-\frac{J_{0}}{N}\sum_{(i,j)}S_{i}S_{j}-\sum_{(i,j)}J_{ij}S_{i}S_{j}
  -\sum_{i}h_{i}S_{i} +D\sum_{i=1}^{N}S_{i}^{2}\,,
  \label{model}
\end{equation}
where the spins can assume values $S=\pm 1,0$.
In the first term, $J_{0}$ represents the uniform ferromagnetic
interaction while $D$ is the anisotropic crystal field.
The disordered interaction $J_{ij}$ is given by
\begin{equation}
  J_{ij}=\frac{J}{N}\big[\xi_{i}\eta_{j}+\xi_{j}\eta_{i}\big]\,,
  \label{Jij}
\end{equation}
in which $\xi_{i}$ and $\eta_{i}$ are independents random variables
following the bimodal distribution
\begin{equation}
  P(x_{i})=\frac{1}{2}\big[\delta(x_{i}-1)+\delta(x_{i}+1)\big]\,.
  \label{pbxi}
\end{equation}
%
The random field $h_{i}$ is also distributed according to the bimodal
distribution:
\begin{equation}
  P(h_{i})= \frac{1}{2} \delta (h_{i}+h_{0}) + \frac{1}{2} \delta (h_{i}-h_{0})\,,
  \label{pbhi}
\end{equation}

The partition function can be written as
\begin{align}
\label{part}
  Z_{N} & = Tr \exp\Bigg[\frac{\beta J_{0}}{N}
    \Bigg(\Big(\sum_{i}S_{i} \Big)^{2} -
    \sum_{i}S^{2}_{i}\Bigg)+\frac{\beta J}{N}
    \Bigg\{\Big(\sum_{i}(\xi_{i}+\eta_{i})S_{i}\Big)^{2} \\ &
    -\Big(\sum_{i} \xi_{i}S_{i}\Big)^{2} - \Big(\sum_{i}\eta_{i}
    S_{i}\Big)^{2} - 2 \sum_{i}\xi_{i}\eta_{i}S_{i}^{2}\Bigg\} + \beta
    h_{i} \sum_{i}S_{i} - \beta D \sum_{i=1}^{N}S_{i}^{2} \Bigg]\,,
  \nonumber 
\end{align}
with $\beta=1/T$ ($T$ is the temperature). The quadratic terms in
Eq. (\ref{part}) can be linearized by the Gaussian identity $\exp
(\lambda a^{2})=\frac{1}{\sqrt{2\pi}}\int^{\infty}_{-\infty}
\left(\frac{-x^{2}}{2}+a\sqrt{2\lambda}x\right)dx$, introducing the SG
order parameters
\begin{equation}
  q_{1}= \frac{1}{N} \sum_{i} \langle\langle\langle \xi_{i}
  S_{i}\rangle\rangle\rangle, ~~~~ q_{2}= \frac{1}{N} \sum_{i}
  \langle\langle\langle \eta_{i} S_{i}\rangle\rangle\rangle.
\end{equation}
and magnetization
\begin{equation}
  m = \frac{1}{N} \sum_{i} \langle\langle\langle
  S_{i}\rangle\rangle\rangle\,.
\end{equation}
At the minimum free energy one has that $q_1=q_2=q$. Thereby, the free
energy per spin is given as
\begin{equation}
  \beta f = \frac{1}{2} \beta J_0 m^2 + \beta J q^2 - \big\langle
  \big\langle\big\langle\ln\big[1+2 e^{-\beta D} \cosh (\beta K)\big]
  \big\rangle\big\rangle\big\rangle\,,
  \label{enli}
\end{equation}
with $K=J_0 m + J(\xi+\eta) q +h$.
The equations for magnetization $m$ and spin glass order parameter $q$
follow from the saddle point equations:
$\frac{\partial}{\partial m } \beta f = 0$ and 
$\frac{\partial}{\partial q } \beta f = 0$,
respectively. Thus
\begin{equation}
  m = \Big\langle\Big\langle\Big\langle \frac{2 e^{-\beta D} \sinh
    (\beta K)} {1+2 e^{-\beta D} \cosh(\beta K)}
  \Big\rangle\Big\rangle\Big\rangle
  \label{mag}
\end{equation}
and
\begin{equation}
  q = \frac{1}{2} \Big\langle\Big\langle\Big\langle (\xi+\eta) \frac{2
    e^{-\beta D}\sinh (\beta K)}{1+2 e^{-\beta D} \cosh(\beta K)}
  \Big\rangle\Big\rangle\Big\rangle\,.
  \label{posg}
\end{equation}
The symbol $\langle\langle\langle ... \rangle\rangle\rangle$
represents the averages on $\xi$, $\eta$ and $h$ which are performed
using the distributions given in Eqs. (\ref{pbxi})-(\ref{pbhi}).

It is also important to investigate the average magnetic occupation.
Therefore, we also calculated $Q=\langle\langle\langle
S_i^2\rangle\rangle\rangle$ which gives
\begin{equation}
  Q = \Big\langle\Big\langle\Big\langle \frac{2 e^{-\beta D} \cosh
    (\beta K)} {1+2 e^{-\beta D} \cosh(\beta K)}
  \Big\rangle\Big\rangle\Big\rangle.
  \label{Q}
\end{equation}
In the context of the present work, $Q$ indicates not only whether the
spin state on the sites are magnetically active to interact with other
spins but also whether they are coupled with the RF. 

\section{Results}

In this section, 
we present and discuss phase diagrams for: (i) $J_0
=0$ which does give $m=0$; (ii) $J_0\neq 0$ which can give $m\neq 0$.
Therefore, in general terms, one can distinguish four phases: (i) SG
with $q\neq 0$ and $m=0$; (ii) ferromagnetic with $q=0$ and $m\neq 0$;
(iii) paramagnetic with $q=0$ and $m=0$ and (iv) the mixed phase with
$q \neq 0$ and $m \neq 0$. As we shall discuss below, these phases are
unfolded into distinct thermodynamic phases with different numerical
values for $q$, $m$ and $Q$.  We also remark that the crystal field
$D$, $h_0$ and $J_0$ are given in units of $J$ (see Eq. (\ref{Jij})).

\subsection{Phase diagrams with $J_0=0$ }
\subsubsection{Temperature $T=0$}

Firstly, it should be remarked that the RF prevents any induced
magnetization.
Therefore, for $J_0=0$,
it is
ruled out any finite induced or spontaneous magnetization.

\begin{figure}
  \centering
  \includegraphics[angle=360,width=8.0cm]{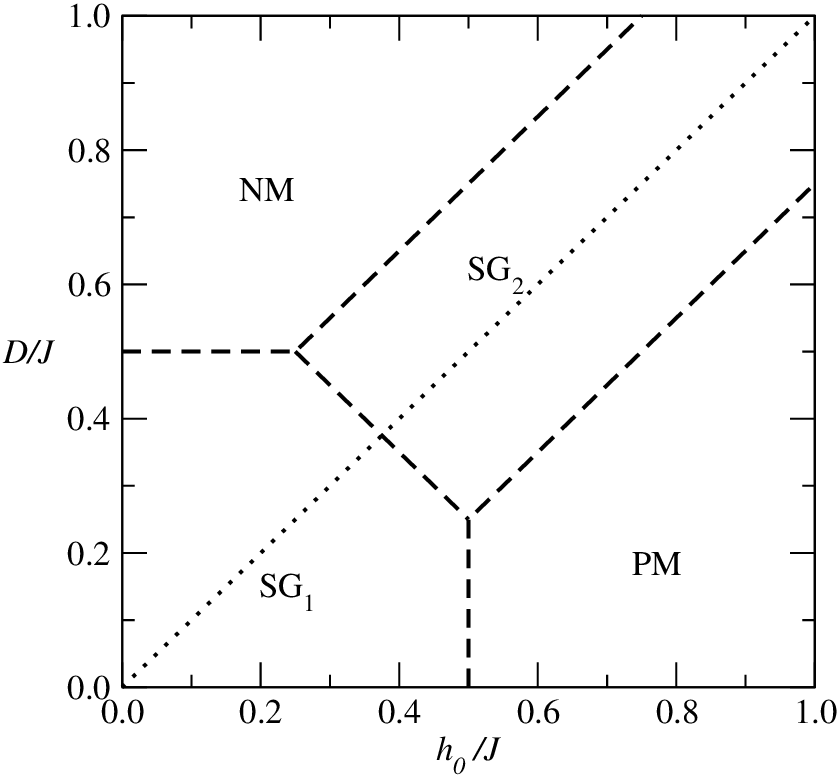}
  \caption{Phase diagram $D/J$ vs $h_0/J$ at $T=0$ with $J_0=0$. The
    order parameters and the average magnetic occupation per site $Q$
    of each phase are given in Table 1. The dashed lines represent
    first order line transitions.  The dotted line represents a
    crossover line separating the behavior of Q in the SG$_1$ and
    SG$_2$ phases for $D\lessgtr h_0$ .}
  \label{Fig1}
\end{figure}

We display in Fig (\ref{Fig1}) the possible ground states in the phase
diagram $D$ vs $h_0$. In the Table I are shown the respective
numerical values of $q$ and $Q$.
There are two regions to be considered initially, large $h_0$ and
small $D$ and vice-versa.  In the first region, for large $D/J$ and
small $h_0/J$, it is found the NM phase while in the second one, for
small $D/J$ and large $h_0/J$, it is found the PM phase.  The NM and
PM phases have $Q=0$ and $1$, respectively. In the region NM, since
$Q=0$, most of the spin states in the sites are magnetically
non-actives preventing, therefore, any spin glass ordering.  In
contrast, in the PM phase most of the spin states are magnetically
active.
However, there is no spin glass solution which indicates that for
$Q=1$,
the RF also prevents any spin glass ordering.
Thus, both situations have $q=0$ but with distinct values for $Q$.
The situation is utterly changed for small $h_0/J$ and $D/J$
($D/J+h_0/J<3/4$, $D/J<1/2$ and $h_0/J<1/2$). There, one can expect
the existence of the usual spin glass ordering found in the AvHSG
model (SG$_1$ in the Fig. (\ref{Fig1}) with $q=1/2$) since there are
conditions for the dominance of a spin glass ordering
provided by the random interaction of the AvHSG model.  Much less
obvious is the existence of a second spin glass ordering (SG$_2$ in
the Fig.  (\ref{Fig1}) with $q=1/4$). The mechanism for the onset of
SG$_2$ phase is uncommon.  For a certain combination of large
$h_0/J$ and $D/J$ ($h_0/J-1/4<D/J<h_0/J+1/4$ and $D/J+h_0/J>3/4$), the
dilution of magnetically active sites favored by $D/J$ competes with
the disordered site activation provided by the increase of $h_0/J$ and
creates again the dominance of a spin glass ordering.
Therefore, the SG phase is unfolded in two distinct phases
SG$_1$ and SG$_2$.

We remark another important difference between SG$_1$ and SG$_2$
phases as compared with NM and PM phases. For instance, in the NM and
PM phases, the magnetic occupation $Q$ does not depend on $h_0$.  In
contrast, inside the SG$_1$ and SG$_2$ phases, the behavior of $Q$ is
determined by the relationship $D\lessgtr h_0$ as can be seen in Table
1. As consequence, the dotted line in Fig. (\ref{Fig1}) represents a
crossover line separating regions with distinct values of $Q$ inside
the same phase (SG$_1$ or SG$_2$) which reflects a fine balance which
determines distinct amounts of magnetically active sites inside SG$_1$
and SG$_2$ phases above and below the line $D=h_0$.  It should be
mentioned that the boundaries lines of the four phases discussed above
Fig. (\ref{Fig1}) are first order ones. As consequence, there are two
triple points in the phase diagram.

\begin{table}[ht]
 \centering
 \caption {Phases, Order Parameters and $Q=\langle S_i^2\rangle$}
 \vspace{0.5cm}
\begin{tabular}{l|rccl} 
 
 Phases & $m$ & $q$ & $Q$ & \\ \hline PM & 0 & 0 & 1\\ SG$_1$ & 0 &
 1/2 & 1 & ($D<h_0$) \\ SG$_1$ & 0 & 1/2 & 1/2 & ($D>h_0$) \\ SG$_2$ &
 0 & 1/4 & 3/4 &($D<h_0$) \\ SG$_2$ & 0 & 1/4 & 1/4 &($D>h_0$) \\ NM &
 0 & 0 & 0 &
\end{tabular} 
\end{table}

\subsubsection{Temperature $T\neq0$}
\label{Tneq0}

\begin{figure}
  \centering \includegraphics[angle=360,width=8.0cm]{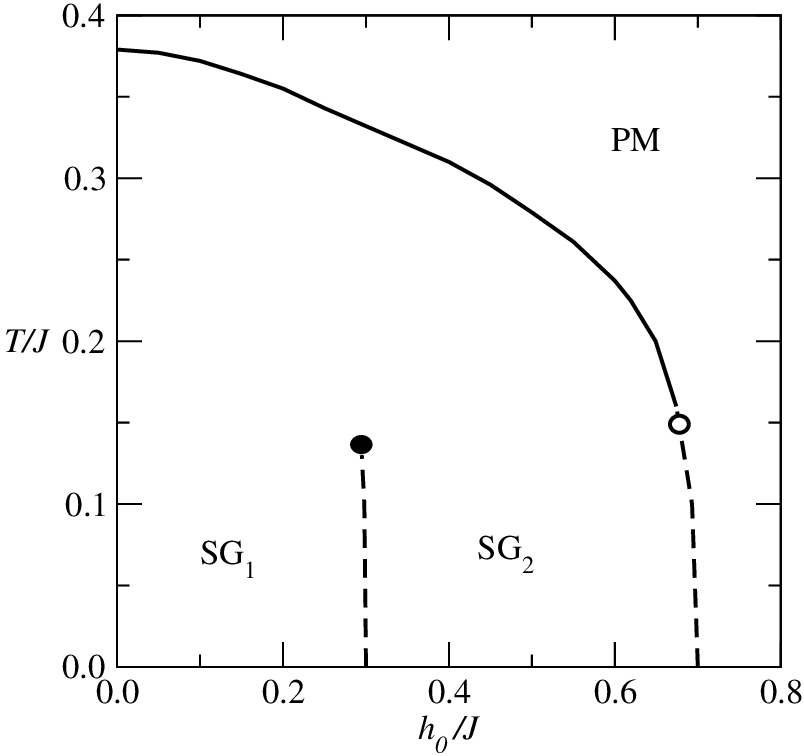}
  \caption{Phase diagram $T/J$ vs $h_0/J$ for $D=0.45J$ presenting one
    ordered critical point (filled circle) and one tricritical point
    (open circle) The dashed lines represent first order line
    transitions.}
  \label{Fig2}
\end{figure}

\begin{figure}
  \centering \includegraphics[angle=360,width=8.0cm]{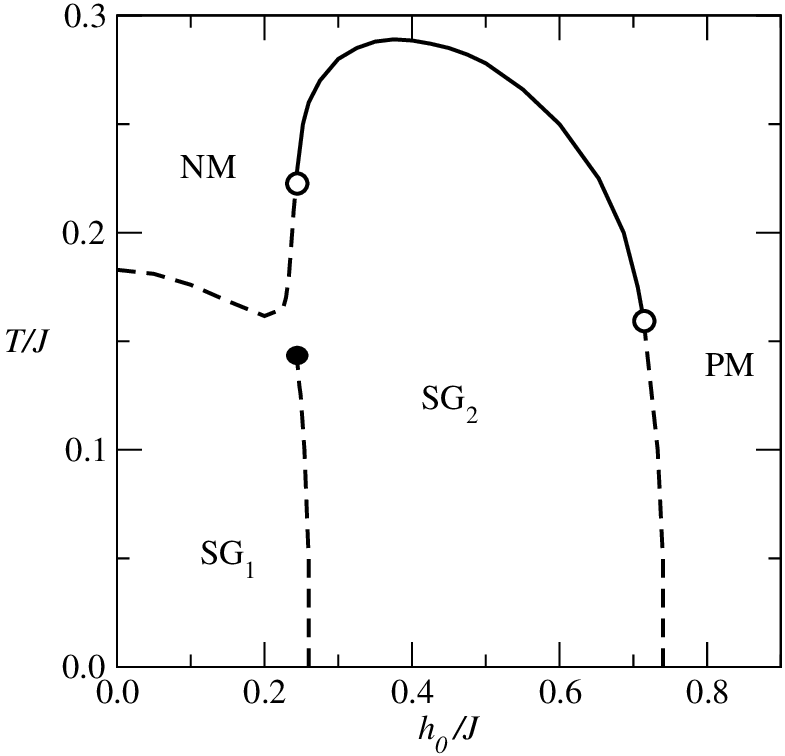}
  \caption{Phase diagram $T/J$ vs $h_0/J$ for $D=0.49J$ presenting one
    ordered critical point (filled circle) and two tricritical points
    (open circles). The dashed lines represent first order line
    transitions.}
  \label{Fig3}
\end{figure}

\begin{figure}
  \centering
  \includegraphics[angle=360,width=8.0cm]{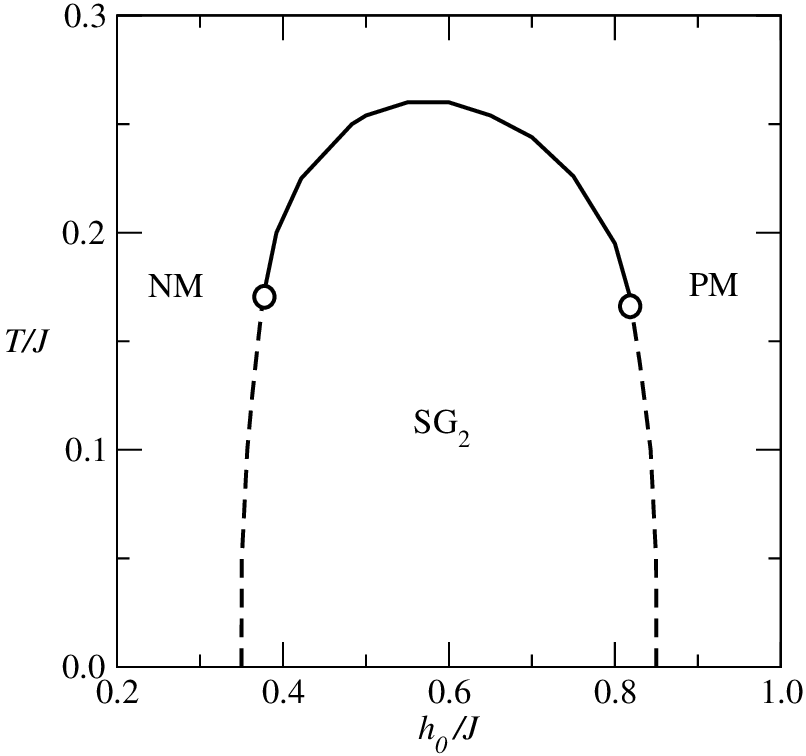}
  \caption{Phase diagram $T/J$ vs $h_0/J$ for $D=0.6J$ presenting two
    tricritical points (open circles).  The dashed lines represent
    first order line transitions.}
  \label{Fig4}
\end{figure}

The effects of temperature increase the complexity of the problem
bringing the presence of tricritical and ordered critical points.  The
sequence of phase diagrams displayed in the
Figs. (\ref{Fig2})-(\ref{Fig4}) has constant $D/J= 0.45$, $0.49$ and
$0.6$, respectively.  We point out that the phase diagrams have
topology similar with those ones found in Blume-Capel model with a RF
given in Ref. \cite{Kaufman1990}.

The values of $D/J$ chosen above select three distinct types which can
be organized in terms of the presence of tricritical and ordered
critical points.  It is well known that, at mean field level, one can
find in a 3-state spins models the presence of tricritical points
which can have two sources: (i) due to the favoring of $S=0$ states at
lower temperature as $D$ increases (for instance, a the Blume-Capel
model \cite{bc}) and ; (ii) reminiscent of Ising limit ($D\rightarrow
-\infty$) of the AvHSG model with bimodal RF. We named the tricritical
points related with the scenario discussed in (i) and (ii) as TC$_1$
and TC$_2$, respectively.
  
The first type of phase diagram can be seen in Fig. (\ref{Fig2}).  At
lower temperature, SG$_1$ and SG$_2$ phases are separated by a first
order line transition ending at an ordered critical point. Then,
SG$_1$ and SG$_2$ phases become identical. Consequently, there is a
second order phase line transition separating the PM phase from the
spin glass phase. This second order line becomes a first order one at
the TC$_2$ point.  The second type of phase diagram is displayed in
Fig.  (\ref{Fig3}).  Besides the presence of an ordered critical point
and the TC$_2$ point, there is also the emergence of the TC$_1$ point
for smaller $h_0/J$.  Other interesting aspect is that the
paramagnetic phase is splitted into NM and PM phases.  The presence of
NM phase for smaller $h_0/J$ suggests that the TC$_1$ point can be
related with the phase transition between the NM (instead of PM) and
any spin glass phase. This relationship is well illustrated in Fig.
(\ref{Fig4}). In the phase diagram shown in that figure, the only
remaining ordered phase is SG$_2$ existing in a dome-shaped
region. Thus, there is no longer an ordered critical point.  The phase
transition line between the SG$_2$ phase and NM for smaller $h_0/J$
and, then, PM phases for larger $h_0/J$ present TC$_1$ and TC$_2$
points. Their location were also checked by the expansion of free
energy in terms of the SG order parameter in the Appendix.

\subsection{Phase diagrams with $J_0 \neq 0$}
\subsubsection{Temperature $T=0$}
\label{Teq0}

The presence of ferromagnetic interaction $J_0\neq 0$ 
brings the possibility of finite spontaneous magnetization and, as
consequence, the arising of mixed or ferromagnetic phases.  In order
to find such phases, we choose $J_0/J=1/2$ and $1$.

\begin{figure}
  \centering
  \includegraphics[angle=0,width=8.0cm]{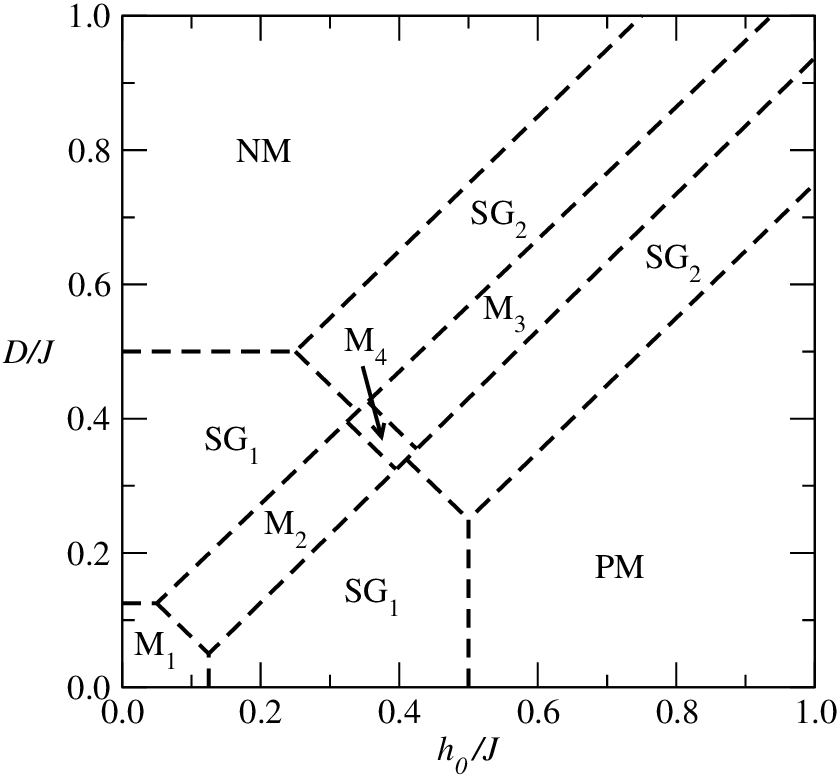}
  \caption{Phase diagram $D/J$ vs $h_0/J$ at $T=0$ with
    $J_0/J=0.5$. The order parameters and the average magnetic
    occupation per site $Q$ of each phase are given in Table II. The
    dashed lines represent first order line transitions.}
  \label{Fig5}
\end{figure}

In the case $J_0/J=1/2$, Table II provides the numerical values for
the $m$, $q$ and $Q$ corresponding to each phase in
Fig. (\ref{Fig5}). The phases SG$_1$, SG$_2$, NM and PM have the same
numerical values for these quantities already displayed in Table I.
In addition, there is the emergence of mixed phases.  It is well
established the existence of a mixed phase in the AvHSG model without
a RF. Nevertheless, our results show that not only part of the spin
glass phase, which is already unfolded, is replaced by a mixed one but
also the mixed phase is unfolded in four distinct phases (M$_1$,
M$_2$, M$_3$ and M$_4$ in Fig.  (\ref{Fig5})) with first order
transition separating them and the SG$_1$ and SG$_2$ phases.  Each
one, M$_1$, M$_2$, M$_3$ or M$_4$ have a different combination of
numerical values for $q$, $m$ and $Q$ (see Table II).  For instance,
inside of the SG$_1$ phase there is the emergence of the M$_1$ phase
for very small values of $D/J$ and $h_0/J$ ($D/J+h_0/J<7/40$,
$D/J<5/40$ and $h_0<5/40$) which corresponds to the usual mixed phase
of the AvHSG model ($m=1/2$, $q=1/2$ and $Q=1$).  However, the
emergence of M$_2$, M$_3$ and M$_4$ is again not obvious.  For these
mixed phases, there is a fine balance of mechanisms driven by the
energy scales $h_0$, $D$ and $J_0$ (given in units of $J$) within the
region where the phases SG$_1$ and SG$_2$ are initially located.
Then, inside the SG$_1$ phase appears also the M$_2$ phase
($7/40<D/J+h_0/J<18/25$ and $-3/40<D/J - h_0/J<3/40$). Inside the
SG$_2$ appears the M$_3$ phase ($D/J+h_0/J>18/25$ and $-3/40<D/J -
h_0/J<3/40$). The M$_4$ phase appears between M$_2$ and M$_3$ phases
($18/25<D/J+h_0/J<39/50$ and $-3/40<D/J - h_0/J<3/40$). It should be
also stressed that this fine balance affects also the average magnetic
occupation per site with the sequence of jumps as can be seen in Table
2 from M$_1$ to M$_4$.  It should be mentioned that due to the number
of phases with first order transitions between them, one has 8 triple
points in Fig. (\ref{Fig5}).  Another aspect to be mentioned is that
for some fixed values of $D$, by varying $h_0$ (or vice-versa), the
SG$_{1}$ and SG$_2$ phases are reentrant.

In the case $J_0/J=1$ there is no longer spin glass or mixed phases
which are entirely replaced by a ferromagnetic one.  This phase is
also unfolded depending on $D/J$ and $h_0/J$. The corresponding FE$_1$
and FE$_2$ phases have $m$ and $Q$ given in Table III.  That case
recovers entirely the results obtained for the Blume-Capel model with
RF given in Ref.  \cite{Kaufman1990}.

\begin{table}[ht]
 \centering
 \caption {Phases, Order Parameters and $Q=\langle S_i^2\rangle$}
 \vspace{0.5cm}
\begin{tabular}{l|rccl} 
 
 Phases  & $m$ & $q$ & $Q$ &\\
 \hline
 PM   & 0   &   0   &  1\\
 M$_1$  & 1/2 &  1/2 &   1 &\\
 M$_2$  & 1/4 &  1/2 &   3/4 &\\
 M$_3$  & 1/4 &  1/4 &   1/2 &\\
 M$_4$  & 3/8 &  3/8 &   5/8 &\\
 SG$_1$ & 0   &  1/2  &  1  & ($D<h_0$) \\
 SG$_1$ & 0   &  1/2  & 1/2 & ($D>h_0$) \\
 SG$_2$ & 0   &  1/4  &  3/4 &($D<h_0$) \\
 SG$_2$ & 0   &  1/4  &  1/4 &($D>h_0$) \\
 NM  & 0   &   0   &  0 &

\end{tabular} 
\end{table}

\begin{table}[ht]
 \centering
 \caption {Phases, Order Parameters and $Q=\langle S_i^2\rangle$}
 \vspace{0.5cm}
\begin{tabular}{l|rccl} 
 
 Phases  & $m$ & $q$ & $Q$ &\\
 \hline
 FE$_1$ & 1   &  0   &   1  & \\
 FE$_2$ & 1/2 &  0   &   1/2 &\\
 NM  & 0   &   0   &  0 & \\
 PM   & 0   &   0   &  1

\end{tabular} 
\end{table}

\subsubsection{Temperature $T\neq0$}

\begin{figure}
  \centering
  \includegraphics[angle=0,width=7.0cm]{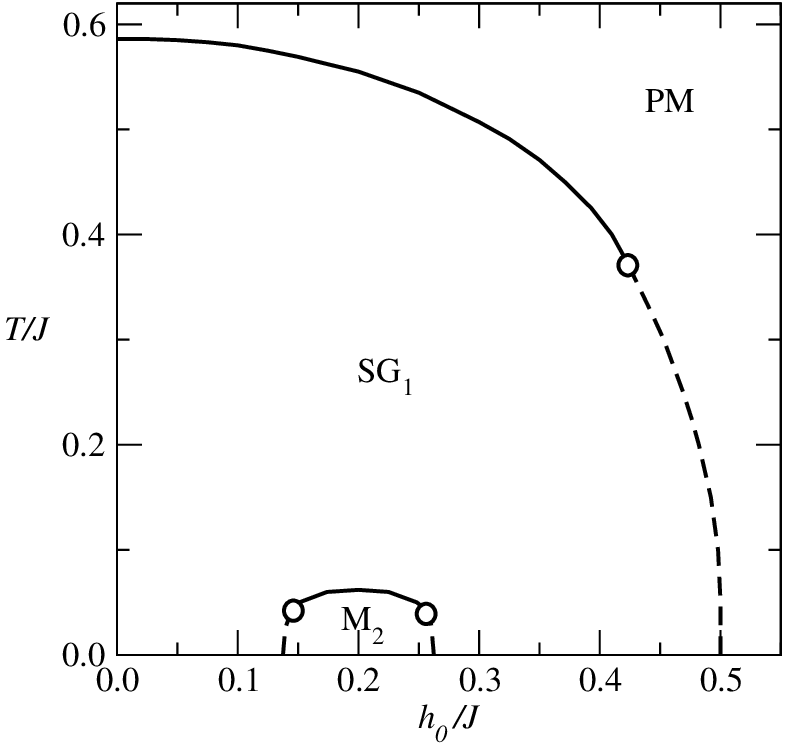}
  \caption{Phase diagram $T/J$ vs $h_0/J$ for $D=0.2J$ presenting
    three tricritical points (open circles).  The dashed lines
    represent first order line transitions.}
  \label{Fig7}
\end{figure}

\begin{figure}
  \centering
  \includegraphics[angle=0,width=7.0cm]{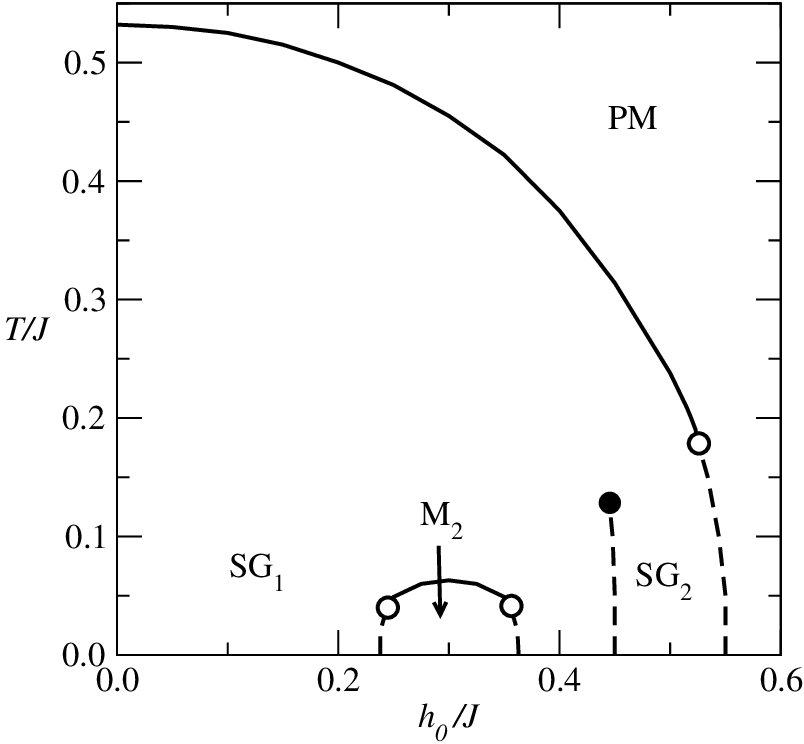}
  \caption{Phase diagram $T/J$ vs $h_0/J$ for $D=0.3J$ presenting one
    ordered critical point (filled circle) and three tricritical
    points (open circles). The dashed lines represent first order line
    transitions.}
  \label{Fig8}
\end{figure}

In this section we discuss only the case $J_0/J=0.5$, since the case
$J_0/J=1$ reproduces exactly the results obtained in
Ref. \cite{Kaufman1990} at finite temperature.

\begin{figure}
  \centering
  \includegraphics[angle=0,width=7.0cm]{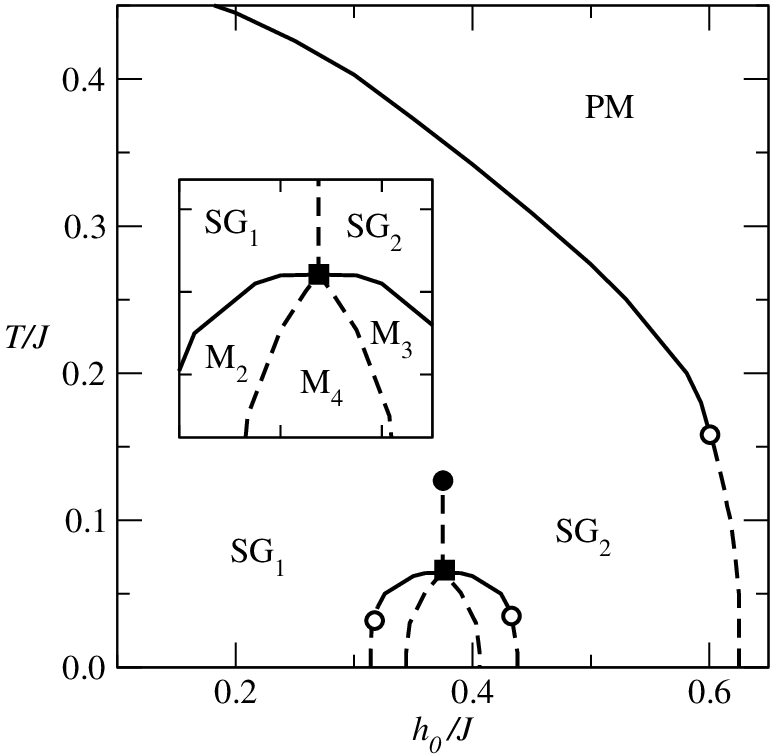}
  \caption{Phase diagram $T/J$ vs $h_0/J$ for $D=0.375J$ presenting
    one ordered critical point (filled circle), one triple point over
    a second order line (filled square) and three tricritical points
    (open circles). The dashed lines represent first order line
    transitions. For a better visualization, the inset shows a
    ``zoom'' of the region with the triple point over the second-order
    line.}
  \label{Fig9}
\end{figure}

\begin{figure}
  \centering
  \includegraphics[angle=0,width=7.0cm]{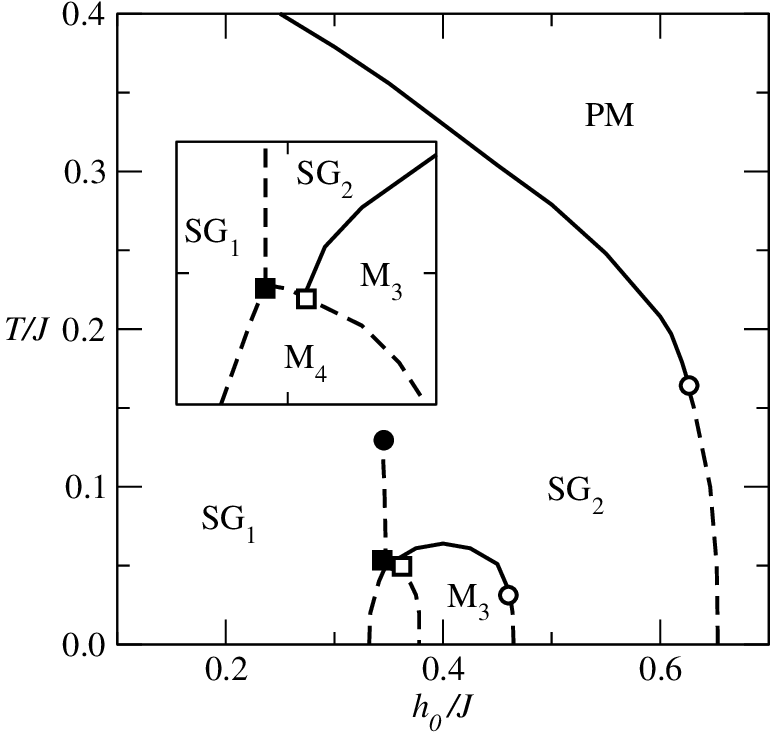}
  \caption{Phase diagram $T/J$ vs $h_0/J$ for $D=0.403J$ presenting
    one ordered critical point (filled circle), one triple point
    (filled square), one critical endpoint (open square) and two
    tricritical points (open circles). The dashed lines represent
    first order line transitions. For a better visualization, the
    inset shows a ``zoom'' of the region with the triple point and the
    critical endpoint. }
  \label{Fig91}
\end{figure}

\begin{figure}
  \centering
  \includegraphics[angle=0,width=7.0cm]{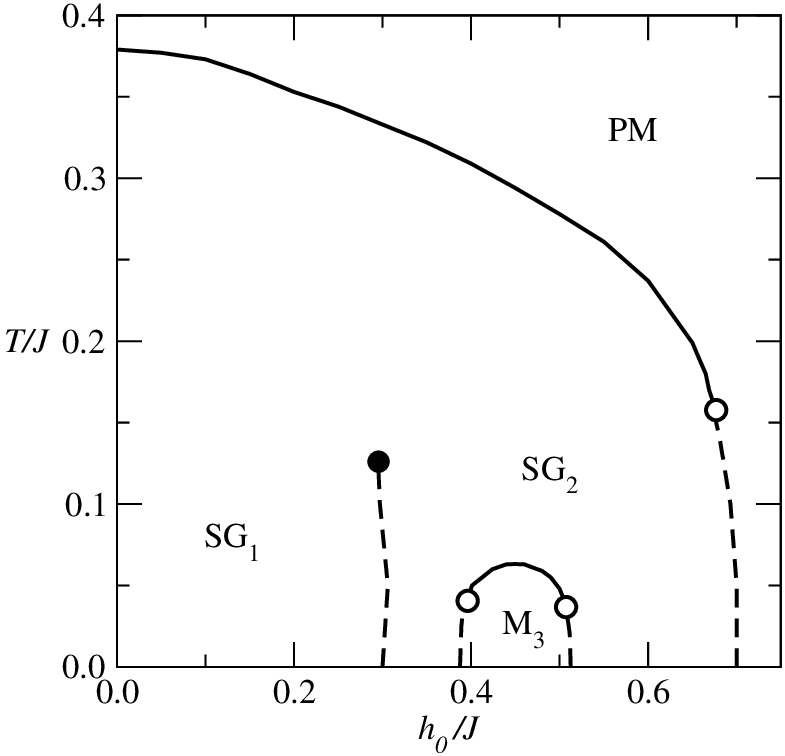}
  \caption{Phase diagram $T/J$ vs $h_0/J$ for $D=0.45J$ presenting one
    ordered critical point (filled circle) and three tricritical
    points (open circles). The dashed lines represent first order line
    transitions.}
  \label{Fig10}
\end{figure}

\begin{figure}
  \centering
  \includegraphics[angle=0,width=7.0cm]{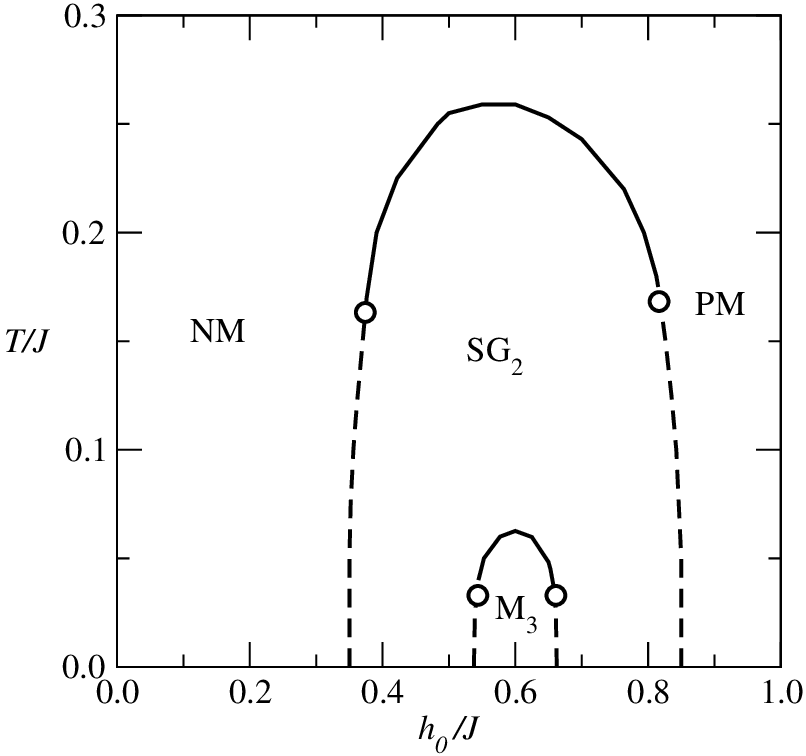}
  \caption{Phase diagram $T/J$ vs $h_0/J$ for $D=0.6J$ presenting four
    tricritical points (open circles). The dashed lines represent
    first order line transitions.}
  \label{Fig11}
\end{figure}

In Figs (\ref{Fig7})-(\ref{Fig11}) are shown phase diagrams $T/J$ vs
$h_0/J$ with $D/J= 0.2$, $0.3$, $0.375$, $0.403$, $0.45$ and $0.6$,
respectively. These choices for values of $D$ allow to display the
most interesting types of phase diagrams in terms of multicritical
points.  These phase diagrams are even more complex as those ones
displayed in Section \ref{Teq0} since the number and the variety of
multicritical points are enhanced.

As a whole, the features observed in Fig. \ref{Fig7} are preserved in
Figs. (\ref{Fig8} - \ref{Fig11}), but there are additional features
that deserve discussion. These figures are representative examples of
$T/J$ vs. $h_0/J$ phase diagrams that are observed in the interval
$1/4<D/J<1/2$, where phases SG$_1$ and SG$_2$ coexist at low
temperature. The first-order line between them ends in a critical
point. It moves in the diagram from the right, like in
Fig. (\ref{Fig8}), to the left, like in Fig. (\ref{Fig10}), as $D/J$
increases. Figure (\ref{Fig8}) corresponds to $D=0.3$. The dome shaped
region is a M$_2$ phase, at the left of the first-order line. Figure
(\ref{Fig10}) corresponds to $D=0.45$. The dome shaped region is a
M$_3$ phase, at the right of the first-order line. Figures
(\ref{Fig9}) and (\ref{Fig91}), for $D/J=0.375$ and $D/J=0.403$,
respectively, show the interesting situation where the first-order
line goes through the dome shaped region, giving rise to the
appearance of the M$_4$ subphase. In Fig. (\ref{Fig9}) M$_4$ appears
between M$_2$ and M$_3$.
Consequently, this figure shows a triple point intersecting two
second-order lines. The triple point joins first-order lines that
separate phases with three distinct $q$, values. These are: $q=0.5$ in
phases M$_2$ and SG$_1$; $q=0.375$ in phase M$_4$; $q=0.25$ in phases
M$_3$ and SG$_2$. The second order line separates the continuous
transition from $m>0$, in phases M$_2$ and M$_3$ to $m=0$, in phases
SG$_1$ and SG$_2$. In Fig. (\ref{Fig91}) M$_2$ is absent, and the
phase diagram shows a triple point between SG$_1$, SG$_2$ and M$_4$
and a critical endpoint where M$_4$ makes first-order transitions with
SG$_2$ and M$_3$ and M$_3$ makes a second-order transition to SG$_2$.

In Fig. (\ref{Fig11}), the situation is completely changed. According
to Fig. (\ref{Fig5}), there SG$_1$ phase is no longer present at
$T/J=0$ if $D/J>1.2$ and, consequently it should not be found at
finite $T$. There is only the SG$_2$ phase which occupies a
dome-shaped spin glass region.  Similarly to the phase diagram shown
in Fig. (\ref{Fig4}), the paramagnetic phase is splitted in NM
(smaller $h_0/J$) and PM (larger $h_0/J$) ones. Consequently, one has
again TC$_1$ and TC$_2$ points.  Furthermore, at lower temperature
there is a reentrant M$_3$ phase occupying a dome-shaped region inside
the SG$_2$ phase.
In the transition line between SG$_2$ phase to M$_3$ one also has two
tricritical points.

\section{Conclusions}\label{conc}

We studied the 3-state van Hemmen spin glass model in the presence of
a bimodal RF, in a mean-field approximation, in three different
scenarios: i) in the absence of uniform ferromagnetic interaction
($J_0=0$), where only spin glass, paramagnetic and non-magnetic phases
are present; ii) in the presence of a moderate uniform ferromagnetic
interaction ($J_0/J=1/2$) where, in addition to the phases mentioned
above, mixed phases were observed; iii) in the presence of a strong
uniform ferromagnetic interaction ($J_0/J=1$), we found ferromagnetic
phases, besides the paramagnetic and non-magnetic phases.  For this
particular case, the phase diagrams reproduce the results found for
the Blume-Capel model with a random field given in
Ref. \cite{Kaufman1990}. It worths to remember that non-magnetic and
paramagnetic phases differ only in the occupation per site $Q$.

For $J_0=0$, the zero-temperature $D/J$ vs. $h_0/J$ phase diagram
displays the unfolding of SG phase in two distinct phases labeled
SG$_{1}$ and SG$_{2}$.  Such unfolding appears as a fine balance
between the RF (that favors active states $S_i=\pm 1$) and the crystal
field $D$ (that favors $S_i=0$).  Furthermore, in both SG$_1$ and
SG$_2$ the occupation per site changes discontinuously over the
diagonal $D/J=h_0/J$ at $T/J=0$.  For finite temperature, besides one
or two (for the dome shaped SG region) tricritical points, there is
also the emergence of a critical point related to the unfolding of the
SG phase.

Far more interesting is the situation for $J_0=1/2$, with the
appearing of several mixed phases and an unusual scenario of triple points and 
multicritical points. The zero
temperature $D/J$ vs. $h_0/J$ phase diagram shows four mixed phases,
labeled M$_1$ to M$_4$, located near the diagonal $D/J=h_0/J$, inside
the two spin glass phases. 
By fixing $D/J$ we obtain several
$T/J$ vs. $h_0/J$ phase diagrams, where the mixed phases appear as a
dome shaped region inside a spin glass phase. The most intriguing  
ones 
are
those 
where the mixed phase dome interacts with the first-order line between
SG$_1$ and SG$_2$, at low temperature, giving rise or to a triple
point intersecting two critical lines
or  a triple point
close to a critical endpoint.

Next, we briefly discuss the contribution of our work as compared with previous 
studies 
of magnetic models with RF. In fact, the results obtained in  the Blume-Capel model 
\cite{Kaufman1990}  or in the Ising van Hemmen model \cite{Nogueira} both with RF  can be 
considered  limit situations of our model. The first case 
corresponds to the situation where $J_0>>J$ while the second one corresponds to 
$D\rightarrow 
-\infty$. Moreover,   the AvHSG model with 
uniform field \cite{Almeida1986} is also contained in our model if the RF distribution in 
Eq. 
\ref{pbhi} 
is generalized as $P(h_{i})= (1-p) \delta (h_{i}+h_{0}) + p \delta 
(h_{i}-h_{0})$ taking the limit $p\rightarrow 1$.

We enphasize that the most important result of our work is to show how 
the combined effects  of $D$ and RF can create conditions to stabilize SG and mixed 
phases. That is far from obvious, 
since the effects of D and RF taken individually tends  
to suppress these phases (at least, at  mean field  level). Particularly, this result 
leads to a complicate scenario of reentrant  phases
when the RF increases. 
Surely, there is a subtil 
mechanism. 
At $T=0$, for small D and RF, it is quite clear that the random interaction is the 
dominant 
energy scale. 
Nevertheless, when RF 
and $D$ increase, they do compete, but mainly to stablish an average magnetic occupation 
which 
is neither $Q=1$ (spins totally active) nor $Q=0$ (spins totally non-active). The main 
consequence is that there are still spins available ($0<Q<1 $) to 
interact via random interaction and, eventually, via ferromagnetic interaction as well. 
That is the reason because SG  and the mixed phases  
stabilise but with order parameter smaller than the corresponding values with small
RF and $D$.
We believe that this mechanism is independent of any specific choice of random 
field 
distribution. It should be remarked that there is one 
common aspect of our result with those ones found in the Ref. \cite{Kaufman1990}. That 
is the
unfolding of phases as well as the proliferation of multicritical points. As demonstrated 
in our work, we can recover entirely the results of Ref. \cite{Kaufman1990}   
when the ferromagnetic interaction  
overcomes the random one.  Indeed, what we are demonstrating is how those
results can develop as long as disordered interaction is being added.

Although our results refer a particular model, we expect that our results can 
shed light in problems containing the interplay  among 
random interaction and field, ferromagnetic interaction and crystal 
field. 
A possible physical realization of such 
scenario might be found, for instance,  in martensitic alloys. In these systems,  
it has been 
stablished the existence of strain glasses which is the analogue of SG concerning lattice 
distortions.  In this glassy phase, 
there are local random configurations of 
lattice distortion \cite{Sherrington2014,Vasseur2010}.   
Finally, we would like to remark that our study does not aim to be exhaustive. So, it is 
not excluded the
possibility for more complicated features, in terms of multicritical
points, for different values of $D$ or $J_0$. On other hand, we remark that we 
are doing
a mean field theory. The existence of some multicritical points may be related with this 
level of description (see, for instance, \cite{Rubem2017,Doria2016}).

\section*{Acknowledgments}

The present study was supported by the brazilian agencies Conselho Nacional de
Desenvolvimento Cient\'{\i}fico e Tecnol\'ogico (CNPq) and CAPES.  The
authors acknowledge Profs. M. C. Barbosa and J. J. Arenzon for useful discussions.

\section{Appendix} 

The tricritical points TC$_1$ and TC$_2$ are obtained from the
expansion of the free energy (see Eq. (\ref{enli})). Therefore:
\begin{equation}  f - f_{0}=  \frac{ J
a_{2}}{K_{0}^{2}}q^{2}+\frac{2 \beta^{3} J^{4} a_{4}}{3
K_{0}^{4}}q^{4} + \frac{4 \beta^{5} J^{6} a_{6}}{45
K_{0}^{6}}q^{6}
\end{equation}
where $\beta f_{0}=-\ln[1+2 e^{-\beta D} \cosh (\beta h_{0})]$
and
\begin{equation} \begin{split} a_{2}= e^{2\beta D} + (2 -  \beta J ) 2 e^{\beta D} 
w  + 
(1- \beta J) 4 w^{2}  + 4 \beta J
v^{2}, 
\end{split} 
\end{equation}
\begin{equation}
\begin{split}
a_{4}= e^{3\beta D} w - 12 e^{\beta D} w^{3}
-16 w^{4} \\
- 8 e^{\beta D} (e^{\beta D} v^{2}-2 w v^{2})
+64 w^{2} v^{2} - 48 v^{4}
\end{split}
\end{equation}
\begin{equation}
\begin{split}
a_{6}= -e^{5\beta D} w + 20 e^{4\beta D} w^{2}+ 80 (e^{3\beta D} w^{3}-e^{2\beta D} 
w^{4})\\ 
-560 e^{\beta D} w^{5}-512 w^{6} +32 e^{4\beta D} v^{2}-344 e^{3\beta D} w v^{2}\\
- 676 e^{2\beta D} w^{2} v^{2}+ 2464 e^{\beta D} w^{3} v^{2}+ 4352  w^{4} v^{2}\\
960 e^{2\beta D} v^{4}-1920 e^{\beta D} w v^{4}- 7680 w^{2} v^{4}+3840 
v^{6}
\end{split}
\end{equation}
with $w=\cosh(\beta h_{0})$, $v=\sinh(\beta h_{0})$ and
$K_{0}=e^{\beta D}+ 2 \cosh(\beta h_{0})$.
To obtain the second order line transition,
we use $a_{2}=0$ and $a_{4}>0$. The tricritical points are located when
$a_{2}=0$, $a_{4}=0$ and $a_{6}>0$.


\begin{thebibliography}{00}
\bibitem{Belanger98} D.~P. Belanger, in {\it Spin Glasses and Random
  Fields}, edited by A.~P.  Young (World Scientific, Singapore, 1998),
  p.~251.
\bibitem{Gingras11} M. Gingras, P. Henelius, J. Phys:
  Conf. Ser. {\bf320}, 012001 (2011).
\bibitem{SNA} R. F. Soares, F. D. Nobre and J. R. L. de Almeida,
  Phys. Rev. B {\bf 50}, 6151 (1994)
\bibitem{Parisi80a} G. Parisi, J. Phys. A {\bf 13}, 1101 (1980); {\it
  ibid.} 1887 (1980).
\bibitem{Almeida78} J. R. L. de Almeida, D. J. Thouless, J. Phys. A
  {\bf 11}, 983 (1978).
\bibitem{Mydosh2015} J. A. Mydosh, Rep. Prog. Phys. {\bf 78}, 052501 (2015).  
\bibitem{gs} S.K. Ghatak and D. Sherrington, J. Phys. C {\bf 10}, 3149
  (1977).
\bibitem{Costa94} F. A. da Costa, C. S. O. Yokoi, S. R. Salinas,
  J. Phys. A {\bf 27}, 3365 (1994).
\bibitem{Vaz2012} C. V. Morais, M. J. Lazo, F. M. Zimmer,
  S. G. Magalhaes, Phys. Rev.  E {\bf 85}, 031133 (2012)
\bibitem{Vaz2013} C. V. Morais, M. J. Lazo, F. M. Zimmer,
  S. G. Magalhaes, Physica A {\bf 392}, 1770 (2013)
\bibitem{Morais2016} C. V. Morais, F. M. Zimmer, M. J. Lazo,
  S. G. Magalhaes, F. D.  Nobre, Phys. Rev. B {\bf 93}, 224206
  (2016).
\bibitem{vh1} J. L. van Hemmen, Phys. Rev. Lett. {\bf 49}, 409 (1982).
\bibitem{vh2} J. L. van Hemmen, A.C.D. van Enter, J. Canisius,
  Z. Phys. B - Condensed Matter {\bf 50}, 311 (1983).
\bibitem{Choy1983} T. C. Choy, D. Sherrington, J. Phys. C: Solid State
  Phys. {\bf 17}, 739 (1984).
\bibitem{Almeida1986} J. R. L. de Almeida, F. C. Brady Moreira,
  Z. Phys. B - Condensed Matter {\bf 63}, 365 (1986).
\bibitem{Nogueira}
Y. Nogueira, J. 
R. Viana, J. R. de Souza, Braz. J. Phys. {\bf 37} 331 (2007) 
\bibitem{Morais}
D. N. de Morais, M. Godoy, A. S. de Arruda J. N. da Silva, J. R. 
de 
Souza, J. Magn. Magn. Mater. {\bf 398} 253-258 (2016)
\bibitem{bc} M. Blume, Phys. Rev., {\bf 141}, 517 (1966).  H.W. Capel,
  Physica, {\bf 32}, 966 (1966).
\bibitem{Kaufman1990} M. Kaufman, M. Kanner, Phys. Rev. B {\bf 42},
  2378 (1990).
\bibitem{Rubem2017} R. Erichsen Jr., A. A. Lopes, S. G. Magalhaes,
  Phys. Rev. E {\bf 95}, 062113 (2017).
\bibitem{Doria2016} F. Doria, R. Erichsen Jr., D. Dominguez,
  M. Gonzalez, S. G.  Magalhaes, Physica A {\bf 422}, 58 (2016).
\bibitem{Sherrington2014} D. Sherrington, Phys. Status Solidi, {\bf 251}, 
1967 (20014).
\bibitem{Vasseur2010} R. Vasseur, T. Lookman, Phys. Rev. B, {\bf 81}, 094107 (2010).

\end{thebibliography}
\end{document}